\documentclass[a4paper]{jpconf}
\usepackage{graphicx}
\usepackage{subfigure}
\begin{document}

\def\nuc#1#2{${}^{#1}$#2}
\def\BBz{0$\nu\beta\beta$}
\def\BBt{2$\nu\beta\beta$}
\def\BB{$\beta\beta$}
\def\Tz{$T^{0\nu}_{1/2}$}
\def\Tt{$T^{2\nu}_{1/2}$}
\def\mj{M{\sc ajo\-ra\-na}}
\def\dem{D{\sc emonstrator}}
\def\mg{M{\sc a}G{\sc e}}
\def\QBB{Q$_{\beta\beta}$}
\def\mBB{$\left < \mbox{m}_{\beta\beta} \right >$}
\def\ge{$^{76}$Ge}

\title{Background model for the \mj\ \dem}

\author{C. Cuesta on behalf of the \mj\ Collaboration\footnote{Collaboration list in~\cite{Nu16Elliott}}}

\address{Center for Experimental Nuclear Physics and Astrophysics, and Department of Physics, University of Washington, Seattle, WA, USA}

\ead{ccuesta@uw.edu}

\begin{abstract}
The \mj\ Collaboration is constructing a system containing 44~kg of high-purity Ge (HPGe) detectors to demonstrate the feasibility and potential of a future tonne-scale experiment capable of probing the neutrino mass scale to $\sim$15~meV. To realize this, a major goal of the \mj\ \dem\ is to demonstrate a path forward to achieving a background rate at or below 1~count/(ROI-t-y) in the 4~keV region of interest (ROI) around the Q-value at 2039~keV. This goal is pursued through a combination of a significant reduction of radioactive impurities in construction materials and analytical methods for background rejection, for example using powerful pulse shape analysis techniques profiting from the p-type point contact (PPC) HPGe detectors technology. The effectiveness of these methods is assessed using simulations of the different background components whose purity levels are constrained from radioassay measurements. Preliminary background results obtained during the engineering runs of the \dem\ are presented.
\end{abstract}

\section{Introduction}

The~\mj~\dem~\cite{Nu16Elliott,mjd} is an array of HPGe detectors that will search for the neutrinoless double beta (\BBz) decay of \ge. The experiment is composed of 44.8~kg of HPGe detectors with 29.7~kg enriched to 88\% of $^{76}$Ge, which also act as the source of \BBz-decay. A modular instrument composed of two cryostats built from ultra-pure electroformed copper is being constructed. The first, Module~1, houses 16.8~kg of enriched germanium detectors and 5.7~kg of natural germanium detectors. Module~1 was moved into the shield, and data taking began, during 2015. The final stage, Module~2 supports 12.8~kg of enriched and 9.4~kg of natural Ge detectors. Module~2 has been assembled and will start taking data soon. The modules are operated in a passive shield that is surrounded by a 4$\pi$ active muon veto and the experiment is placed at 4850-ft depth at the Sanford Underground Research Facility in Lead, SD, USA~\cite{surf}.

\section{Background Model}

\begin{figure}[h]
\begin{center}
\begin{minipage}{16pc}
\begin{center}
\includegraphics[width=1\textwidth]{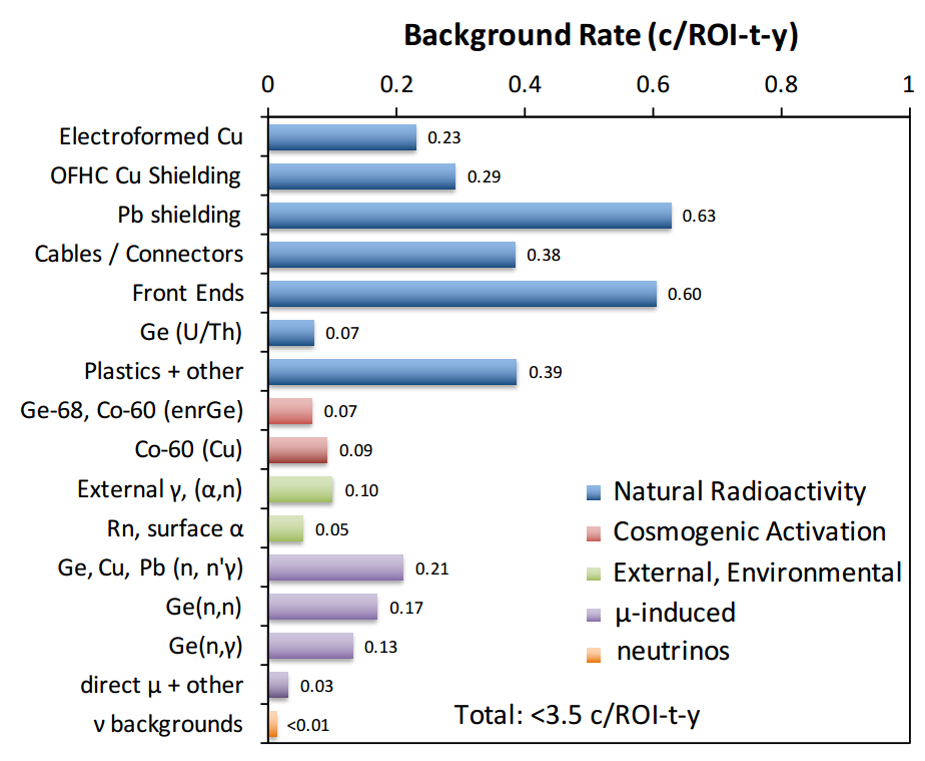}
\caption{\label{fig:assay}Estimated background contributions in the ROI for the \BBz-decay search, measured in counts/(ROI-t-y).}
\end{center}
\end{minipage}\hspace{2pc}%
\begin{minipage}{18pc}
\begin{center}
\includegraphics[width=1\textwidth]{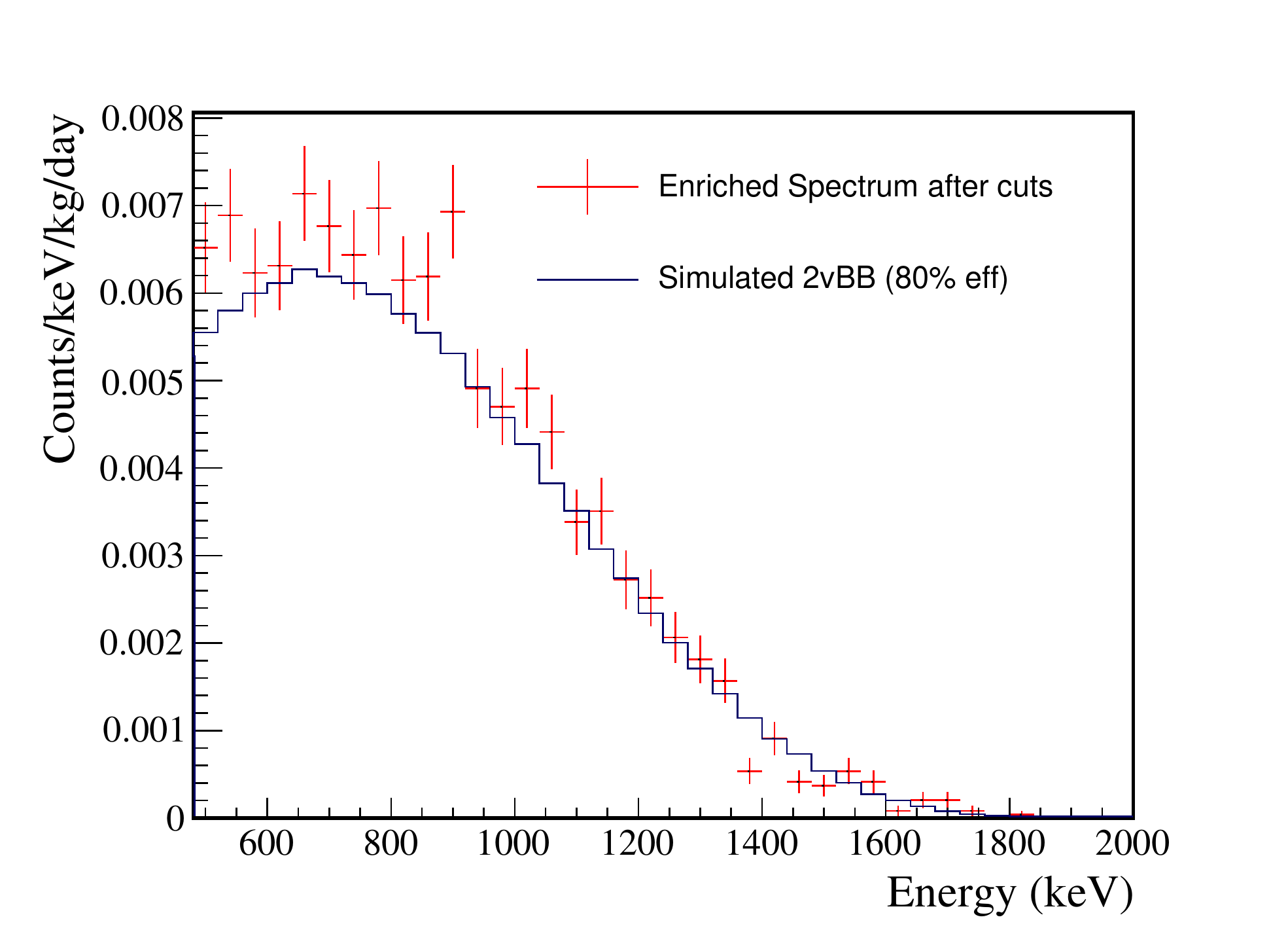}
\caption{\label{fig:2bb}Energy spectrum of $^{enr}$Ge detectors for DS1 after cuts compared to the simulated \BBt\ spectrum.}
\end{center}
\end{minipage}\hspace{2pc}%
\end{center}
\end{figure}

The main technical challenge of the \mj\ \dem\ is to reach a background rate of 3~counts/(ROI-t-y) after analysis cuts, which projects to a background level of 1~count/(ROI-t-y) in a large scale experiment after accounting for additional shielding, better self-shielding, and if necessary, increased depth. To achieve this goal, background sources must be reduced and offline background rejection must be maximized. The estimated background contributions in the ROI, based on achieved assays~\cite{mjdassay} of materials and simulations for cosmic muon interactions, sum to $<$3.5~counts/(ROI-t-y) in the \mj\ \dem, see Fig.~\ref{fig:assay}.

Module~1 has been taking data since June 2015, and data blinding started in April 2016. The data are divided in two data sets: DS0 and DS1. In Fall 2015, we implemented planned improvements: installed the inner copper shield, added additional shielding within the vacuum of the cryostat cross arm, exchanged the cryostat seal for one with lower radioactivity, and repaired non-operating channels. These changes define the difference between DS0 and DS1, and DS1 is being used to determine the background. The following cuts are applied to remove non-plausible \BBz-decay candidate events:

\begin{itemize}
  \item Data reduction: Instrumental background events, like transient noise due to LN fills, are tagged and rejected for the analysis.
  \item Muon related events: Triggered events in the HPGe detectors that were within 1~s after an active muon veto trigger are removed.
  \item Multiplicity: Events with more than one triggered detector are removed. Double beta-decay events occur in a single detector only.
  \item A/E: The sharply peaked weighting potential of a PPC Ge detector results in distinct signal shapes between multiple interaction site events (e.g. from gamma rays) and single-site double beta-decay events. The multi-site events exhibit multiple steps in the charge signal waveform and, correspondingly, multiple peaks in the current signal waveform. However, single-site events give a single peak. A cut based on normalizing the maximum current amplitude by the calibrated energy (A/E) is used to reject multi-site events.
  \item Delayed charge recovery: We identified background in the energy range 1.7-3.5~MeV likely arising from alpha particles impinging on the detectors' passivated surfaces. The slow mobility of electrons on the passivated surface results in a very slow collection of a portion of the charge. This produces a distinctive shape that can be effectively cut.
\end{itemize}

\begin{figure}[h]
\begin{center}
\subfigure{\includegraphics[width=0.52\textwidth]{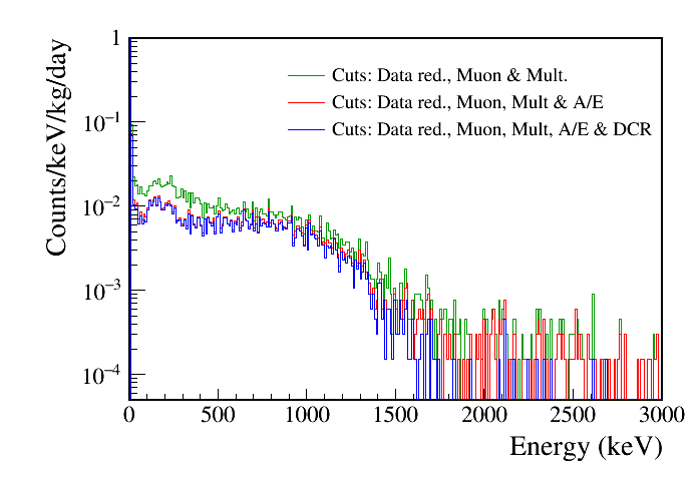}}
\subfigure{\includegraphics[width=0.46\textwidth]{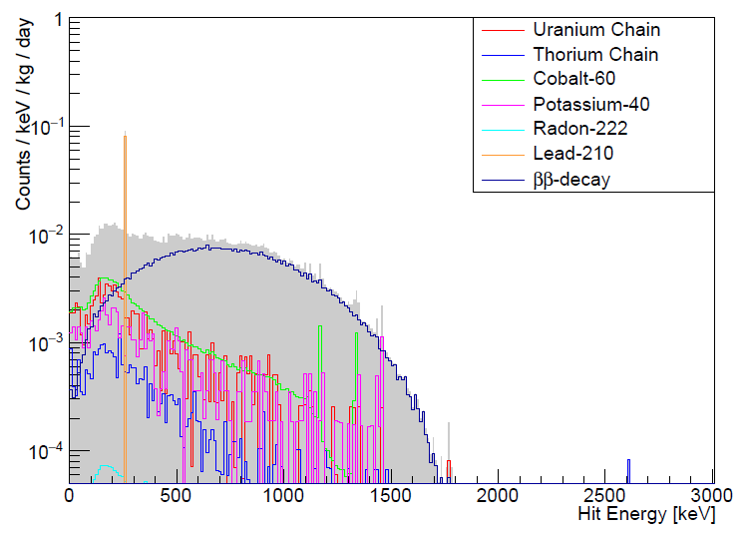}}
\caption{Left, background energy spectra of the HPGe detectors with the different analysis cuts applied for DS1 shown with 10 keV/bin. Efficiency corrections have been applied. It is worth mentioning that A/E and DCR efficiencies are estimated for E\,$>$\,1~MeV. Right, simulated event energy spectrum for all detectors without cuts for DS1 showing the contributions from the different isotopes. Some activities considered are upper limits. The grey spectrum is the sum of all background sources.}
\label{fig:cuts}
\end{center}
\end{figure}

The live time taking into account all these cuts and efficiencies is 46.9~d (53.6~d) for DS0 (DS1) corresponding to 501.4~kg$\cdot$d (606.0~kg$\cdot$d) $^{enr}$Ge exposure. DS0 and DS1 experimental spectra are used to determine the background sources contributing in all energy ranges. As expected, a significant background reduction is observed in DS1. The enriched detectors in DS1 are used to estimate the background. Activities from U, K, and Th are being estimated and preliminary upper limits are $<$0.03 counts/(kg$\cdot$d) per isotope. Cosmogenic contributions are also being evaluated. Monte Carlo simulations are carried out with MaGe~\cite{mage}, a simulation software framework based on Geant4~\cite{geant4} and developed by the \mj\ and GERDA collaborations. Simulations are being used to validate the background model and the pulse-shape analysis techniques, and to study other systematic effects. Figures~\ref{fig:2bb} and~\ref{fig:cuts} show the background energy spectra in DS1 compared to simulations.


\section*{Acknowledgments}
This material is based upon work supported by the U.S. Department of Energy, Office of Science, Office of Nuclear Physics. We acknowledge support from the Particle Astrophysics and Nuclear Physics Programs of the National Science Foundation. This research uses these US DOE Office of Science User Facilities: the National Energy Research Scientific Computing Center and the Oak Ridge Leadership Computing Facility. We acknowledge support from the Russian Foundation for Basic Research. We thank our hosts and colleagues at the Sanford Underground Research Facility for their support.

\section*{References}

\bibliographystyle{iopart-num}
\bibliography{CCuesta_Neutrino16}

\end{document}